\documentclass[10pt,conference]{IEEEtran}
\IEEEoverridecommandlockouts
\usepackage[left=0.7in, right=0.7in, bottom=1.015in, top=0.7in]{geometry}

\makeatletter
\newcommand{\AddInputPath}[1]{%
  \ifx\input@path\@undefined
    \def\input@path{#1}
  \else
    \g@addto@macro{\input@path}{#1}
  \fi
}
\makeatother
\AddInputPath{{../}}

\usepackage{shellesc}

\usepackage{etex}
\usepackage{relsize}

\usepackage[dvipsnames,svgnames,usenames]{xcolor}
\usepackage{array}
\usepackage{booktabs,tabularx}
\usepackage{multirow}
\usepackage[per-mode=symbol,detect-mode=true]{siunitx}
\usepackage{graphicx}

\usepackage[T1]{fontenc}
\usepackage{textcomp}
\usepackage[utf8]{inputenc}
\usepackage[final]{microtype}
\usepackage{icomma}
\usepackage{xspace}

\usepackage[tbtags]{amsmath}
\usepackage{amssymb,amsfonts,bm}
\usepackage{mathtools} 
\usepackage{dsfont}
\usepackage{mathrsfs}
\usepackage{accents}
\usepackage{empheq}
\usepackage{nccmath}
\usepackage{balance}
\usepackage{setspace}

\usepackage{color}
\usepackage{calc}
\usepackage{tikz}
\usepackage{pgfplots,pgfplotstable}
\usetikzlibrary{pgfplots.groupplots}
\usetikzlibrary{matrix} %Tikz table
\pgfplotsset{compat=1.18} 
\usepackage{pdftexcmds}
\makeatletter
\newcommand{\strequal}[2]{\pdf@strcmp{#1}{#2}==0}
\makeatother

\usepackage[capitalize]{cleveref}
\usepackage[font=small]{subcaption}

\usepackage[inline]{enumitem}
\usepackage{algorithm}
\usepackage{algpseudocode}
\makeatletter
\newcommand{\algmargin}{\the\ALG@thistlm}
\makeatother
\newlength{\whilewidth}
\algdef{SE}[parWHILE]{parWhile}{EndparWhile}[1]
  {\parbox[t]{\dimexpr\linewidth-\algmargin}{%
     \hangindent\whilewidth\strut\algorithmicwhile\ #1\ \algorithmicdo\strut}}{\algorithmicend\ \algorithmicwhile}%
\algnewcommand{\parState}[1]{\State%
  \parbox[t]{\dimexpr\linewidth-\algmargin}{\strut #1\strut}}

\makeatletter
\newcommand\fs@spaceruled{\def\@fs@cfont{\bfseries}\let\@fs@capt\floatc@ruled
  \def\@fs@pre{\vspace{.05in}\hrule height.8pt depth0pt \kern2pt}%
  \def\@fs@post{\kern2pt\hrule\relax}%
  \def\@fs@mid{\kern2pt\hrule\kern2pt}%
  \let\@fs@iftopcapt\iftrue}
\makeatother

\usepackage{glossaries}
\usepackage{ifthen}
\usepackage[noadjust]{cite}
\usepackage{multibib}

\usepackage{comment}
\usepackage{todonotes}
\let\legacytodo\todo
\newcommand{\ruggedtodo}[2][]{\tikzexternaldisable\legacytodo[#1]{#2}\tikzexternalenable}
\renewcommand{\todo}[1]{\ruggedtodo[inline]{#1}}
\usetikzlibrary {arrows.meta}

\makeglossaries

\newacronym{fso}{FSO}{free-space optical communication}
\newacronym{tle}{TLE}{two-line element set}
\newacronym{ai}{AI}{artificial intelligence}
\newacronym{ann}{ANN}{artificial neural network}
\newacronym{jscc}{JSCC}{joint source-channel coding}
\newacronym{raan}{RAAN}{right ascension of the ascending node}
\newacronym{uav}{UAV}{unmanned aerial vehicle}
\newacronym{haps}{HAPS}{high-altitude platform station}
\newacronym{6g}{6G}{sixth generation}
\newacronym{cgr}{CGR}{contact graph routing}
\newacronym{dtn}{DTN}{delay-tolerant networking}
\newacronym{fl}{FL}{federated learning}
\newacronym{fo}{FO}{federated optimization}
\newacronym{dl}{DL}{deep learning}
\newacronym{fedavg}{FedAvg}{federated averaging}
\newacronym{dml}{DML}{distributed ML}
\newacronym{ps}{PS}{parameter server}
\newacronym{ml}{ML}{machine learning}
\newacronym{sgd}{SGD}{stochastic gradient descent}
\newacronym{dsgd}{DSGD}{distributed stochastic gradient descent}
\newacronym{isl}{ISL}{inter-satellite link}
\newacronym{gsl}{GSL}{ground-satellite link}
\newacronym{gs}{GS}{ground station}
\newacronym{ecef}{ECEF}{earth-centered, earth-fixed}
\newacronym{eci}{ECI}{Earth-centered inertial}
\newacronym{ofdm}{OFDM}{orthogonal frequency-division multiplexing}
\newacronym{los}{LoS}{line-of-sight}
\newacronym{leo}{LEO}{low earth orbit}
\newacronym{meo}{MEO}{medium earth orbit}
\newacronym{gso}{GSO}{geosynchronous orbit}
\newacronym{geo}{GEO}{geostationary}
\newacronym{ntn}{NTN}{non-terrestrial networks}
\newacronym{eo}{EO}{Earth observation}
\newacronym{iot}{IoT}{Internet of Things}
\newacronym{irs}{IRS}{intelligent reflecting surface}
\newacronym{socp}{SOCP}{second-order cone program}
\newacronym{soc}{SOC}{second-order cone}
\newacronym{dsl}{DSL}{digital subscriber line}
\newacronym{wsee}{WSEE}{weighted sum energy efficiency}
\newacronym{mmwave}{mmWave}{millimeter wave}
\newacronym{dfg}{DFG}{Deutsche Forschungsgemeinschaft}
\newacronym{haec}{HAEC}{Highly Adaptive Energy-Efficient Computing}
\newacronym{hpc}{HPC}{High Performance Computing}
\newacronym{mac}{MAC}{multiple-access channel}
\newacronym{bc}{BC}{broadcast channel}
\newacronym{siso}{SISO}{single-input single-output}
\newacronym{simo}{SIMO}{single-input multiple-output}
\newacronym{miso}{MISO}{multiple-input single-output}
\newacronym{mimo}{MIMO}{multiple-input multiple-output}
\newacronym{af}{AF}{amplify-and-forward}
\newacronym{df}{DF}{decode-and-forward}
\newacronym{cf}{CF}{compress-and-forward}
\newacronym{mwrc}{MWRC}{multi-way relay channel}
\newacronym{dmmwrc}{DM-MWRC}{discrete memoryless multi-way relay channel}
\newacronym{pde}{PDE}{partial data exchange}
\newacronym{fde}{FDE}{full data exchange}
\newacronym{iid}{i.i.d.\@}{independent and identically distributed}
\newacronym{di}{DI} {difference of increasing}
\newacronym{dc}{DC}{difference of convex}
\newacronym{mm}{MM}{mixed monotonic}
\newacronym{mmp}{MMP}{mixed monotonic programming}
\newacronym{awgn}{AWGN}{additive white Gaussian noise}
\newacronym{wgn}{WGN}{white Gaussian noise}
\newacronym{awg}{AWG}{additive white Gaussian}
\newacronym{sic}{SIC}{successive interference cancellation}
\newacronym{snr}{SNR}{signal-to-noise ratio}
\newacronym{sinr}{SINR}{signal to interference plus noise ratio}
\newacronym{inr}{INR}{interference to noise ratio}
\newacronym{zf}{ZF}{zero-forcing}
\newacronym{mrt}{MRT}{maximum ratio transmission}
\newacronym{mmse}{MMSE}{minimum mean square error}
\newacronym{sud}{SUD}{single user decoding}
\newacronym{dof}{DoF}{degrees of freedom}
\newacronym{gdof}{GDoF}{generalized degrees of freedom}
\newacronym{nnc}{NNC}{noisy network coding}
\newacronym{dmn}{DMN}{discrete memoryless network}
\newacronym{csi}{CSI}{channel state information}
\newacronym{pmf}{pmf}{probability mass function}
\newacronym{dmic}{DM-IC}{discrete memoryless interference channel}
\newacronym{ic}{IC}{interference channel}
\newacronym{gic}{GIC}{Gaussian interference channel}
\newacronym{if}{IF}{interference}
\newacronym{ee}{EE}{energy efficiency}
\newacronym{gee}{GEE}{global energy efficiency}
\newacronym{tin}{TIN}{treating interference as noise}
\newacronym{snd}{SND}{simultaneous non-unique decoding}
\newacronym{sd}{SD}{simultaneous decoding}
\newacronym{hk}{HK}{Han-Kobayashi}
\newacronym{rs}{RS}{rate splitting}
\newacronym{rf}{RF}{radio frequency}
\newacronym{pa}{PA}{power amplifier}
\newacronym{lna}{LNA}{low noise amplifier}
\newacronym{lo}{LO}{local oscillator}
\newacronym{adc}{ADC}{analog-to-digital converter}
\newacronym{dac}{DAC}{digital-to-analog converter}
\newacronym{dsp}{DSP}{digital signal processing}
\newacronym{brd}{BRD}{best response dynamics}
\newacronym{br}{BR}{best response}
\newacronym{ne}{NE}{Nash equilibrium}
\newacronym{lhs}{LHS}{left-hand side}
\newacronym{rhs}{RHS}{right-hand side}
\newacronym{ran}{RAN}{radio access network}
\newacronym{qos}{QoS}{Quality of Service}
\newacronym{ngmn}{NGMN}{Next Generation Mobile Networks}
\newacronym{cap}{CAP}{Capacity Adaptation}
\newacronym{bwa}{BW}{Bandwidth Adaptation}
\newacronym{prb}{PRB}{physical resource block}
\newacronym{se}{SE}{spectral efficiency}
\newacronym{tp}{TP}{throughput}
\newacronym{bs}{BS}{base station}
\newacronym{ue}{UE}{user equipment}
\newacronym{mop}{MOP}{multi-objective optimization problem}
\newacronym{gda}{GDA}{generalized Dinkelbach's algorithm}
\newacronym{midcp}{MIDCP}{mixed integer disciplined convex programming}
\newacronym{lp}{LP}{linear program}
\newacronym{brb}{BRB}{branch reduce and bound}
\newacronym{bb}{BB}{branch and bound}
\newacronym{sit}{SIT}{successive incumbent transcending}
\newacronym{oma}{OMA}{orthogonal multiple access}
\newacronym{noma}{NOMA}{non-orthogonal multiple access}
\newacronym{wlog}{w.l.o.g.\@}{without loss of generality}
\newacronym{lsc}{l.s.c.\@}{lower semi-continuous}
\newacronym{usc}{u.s.c.\@}{upper semi-continuous}
\newacronym{kkt}{KKT}{Karush-Kuhn-Tucker}
\newacronym{ptp}{PTP}{point-to-point}
\newacronym{fspl}{FSPL}{Free space path loss}
\newacronym{sfl}{SFL}{satellite federated learning}
\newacronym{gu}{GU}{global update}
\newacronym{cu}{CU}{cluster update}
\glsenableentrycount
\makeglossaries
\usetikzlibrary{positioning}
\usetikzlibrary{calc}
\usetikzlibrary{math}
\usetikzlibrary{fit}
\usetikzlibrary{intersections}
\usetikzlibrary{decorations.pathreplacing}
\usetikzlibrary{decorations.markings}
\usetikzlibrary{3d,angles}
\usetikzlibrary{arrows.meta}

\pgfdeclarelayer{background}
\pgfsetlayers{background,main}

\usetikzlibrary{external}
\usepgfplotslibrary{colorbrewer}

\tikzset{
	small1/.style={fill=DeepPink},
	small2/.style={fill=DeepSkyBlue},
	small3/.style={fill=MediumSpringGreen},
	ps/.style={fill=Gold},
	link/.style = {semithick},
	plane/.style={plane origin={(#1,0,0)}, plane x = {(#1,0,1)}, plane y = {(#1,1,0)}, rotate around y = -9, canvas is plane}
}

\tikzset{
	antenna/.pic={
		\draw[thick] (0,0) -- ++(120:2mm) -- ++(0:2mm) -- cycle -- (0,-1.5mm);
	}
}


\crefname{equation}{}{}
\crefrangeformat{equation}{(#3#1#4)--(#5#2#6)}
\crefmultiformat{equation}{(#2#1#3)}{ and~(#2#1#3)}{, (#2#1#3)}{, (#2#1#3)}
\crefrangemultiformat{equation}{#3(#1)#4--#5(#2)#6}{, #3(#1)#4--#5(#2)#6}{, #3(#1)#4--#5(#2)#6}{, #3(#1)#4--#5(#2)#6}
\undef\mod
\DeclareMathOperator\mod{mod}

\newcommand{\norm}[1]{\ensuremath{\left\lVert #1 \right\rVert}}

\let\vec\bm

\allowdisplaybreaks[3]

\DeclareSIUnit \dBm {dBm}
\DeclareSIUnit \dBW {dBW}
\DeclareSIUnit \dBi {dBi}
\DeclareSIUnit \bpcu {bpcu}

\DeclareFontFamily{U}{mathx}{\hyphenchar\font45}
\DeclareFontShape{U}{mathx}{m}{n}{
      <5> <6> <7> <8> <9> <10>
      <10.95> <12> <14.4> <17.28> <20.74> <24.88>
      mathx10
      }{}
\DeclareSymbolFont{mathx}{U}{mathx}{m}{n}
\DeclareMathSymbol{\bigtimes}{1}{mathx}{"91}

\usepackage{pgfkeys}
    \def\addlegendimage{\csname pgfplots@addlegendimage\endcsname}

\pgfplotscreateplotcyclelist{default}{%
	blue,mark=*\\%
	red,mark=star\\%
	teal,mark=square*\\%
	brown!60!black,mark=otimes*\\%
}

\definecolor{plot1}{RGB}{228,26,28}
\definecolor{plot2}{RGB}{55,126,184}
\definecolor{plot3}{RGB}{77,175,74}
\definecolor{plot4}{RGB}{152,78,163}
\definecolor{plot5}{RGB}{255,127,0}
\definecolor{plot6}{RGB}{166,86,40}
\tikzstyle{fedsatschedule}=[plot1]
\tikzstyle{fedsat}=[plot2]
\tikzstyle{fedisl}=[plot3]
\tikzstyle{fedavg}=[plot4]
\tikzstyle{fedasync1}=[plot5]
\tikzstyle{fedasync2}=[plot6]

\newcolumntype{P}[1]{>{\centering\arraybackslash}p{#1}}

\ifCLASSOPTIONdraftcls
\AtBeginEnvironment{figure}{}
\fi

\AtBeginEnvironment{algorithmic}{\footnotesize}

\begin{document}
\bstctlcite{IEEEexample:BSTcontrol}
\IEEEoverridecommandlockouts\IEEEpubid{\makebox[\columnwidth]{ 978-1-6654-3540-6/22~\copyright~2022 IEEE \hfill} \hspace{\columnsep}\makebox[\columnwidth]{ }}

%\title{Scheduling in Satellite Federated Learning }
\title{Scheduling for On-Board Federated Learning with Satellite Clusters}
%\title{Scheduling in Satellite Federated Learning }
%On-board Federated Learning for Satellite Clusters with Inter-Satellite Links
% Assisted, For, In
{\author{\IEEEauthorblockN{Nasrin Razmi\IEEEauthorrefmark{1}\IEEEauthorrefmark{2}, Bho Matthiesen\IEEEauthorrefmark{1}\IEEEauthorrefmark{2}, Armin Dekorsy\IEEEauthorrefmark{1}\IEEEauthorrefmark{2}, and Petar Popovski\IEEEauthorrefmark{3}\IEEEauthorrefmark{1}}

\IEEEauthorblockA{\IEEEauthorrefmark{1} Dept.\ of Communications Engineering, University of Bremen, Germany}

\IEEEauthorblockA{\IEEEauthorrefmark{2}Gauss-Olbers Space Technology
Transfer Center, University of Bremen, Germany}

\IEEEauthorblockA{\IEEEauthorrefmark{3} Dept. of Electronic Systems, Aalborg University, Denmark}

\IEEEauthorblockA{Emails:  \{razmi, matthiesen, dekorsy\}@ant.uni-bremen.de, petarp@es.aau.dk}

\thanks{This work was funded in part
by the German Research Foundation (DFG) under Germany's Excellence Strategy (EXC 2077 at University of Bremen, University Allowance).}
}}

\maketitle

\begin{abstract}
Mega-constellations of small satellites have evolved into a source of massive amount of valuable data.
To manage this data efficiently, on-board federated learning (FL) enables satellites to train a machine learning (ML) model collaboratively without having to share the raw data. This paper introduces a scheme for scheduling on-board FL for constellations connected with intra-orbit inter-satellite links.
The proposed scheme utilizes the predictable visibility pattern between satellites and ground station (GS), both at the individual satellite level and cumulatively within the entire orbit, to mitigate intermittent connectivity and best use of available time.
To this end, two distinct schedulers are employed: one for coordinating the FL procedures among orbits, and the other for controlling those within each orbit.
These two schedulers cooperatively determine the appropriate time to perform global updates in GS and then allocate suitable duration to satellites within each orbit for local training, proportional to usable time until next global update.
This scheme leads to improved test accuracy within a shorter time.

\end{abstract}


\section{Introduction}

Mega-constellations of low Earth orbit (LEO) satellites play a crucial role in various applications, ranging from global internet coverage and Earth observation to space exploration \cite{9217520, LiuConsMag, luo2022very}.
However, the deluge of data produced by the unprecedented numbers of satellites, e.g., high-resolution hyperspectral images, poses a challenge for transmitting them back to the Earth, particularly considering the limitations on available bandwidth and constraints on delay. 
%downlink, substantial amount
Furthermore, specific types of data, e.g., satellite-captured images obscured by clouds, may not be suitable to be sent to the Earth \cite{giuffrida2020cloudscout}.

% meet these demand?
To tackle these challenges, \cgls{ml} algorithms can be implemented in on-board of satellites to extract valuable insight from the data, leading to reduced needs for communication and improved operational efficiency. %facilitating timely operations
A notable example is the Phisat-1, ESA mission \cite{giuffrida2020cloudscout}, which employs a convolutional neural network (CNN) to transmit only non-cloudy images to the Earth, while discarding those with cloud level exceeding a predefined threshold.

To attain an accurate \cgls{ml} model, exploiting data of all satellites is essential. However, constraints such as limited bandwidth impede the sharing of raw data.  
%and perhaps data privacy
%Here, \cgls{fl} offers a solution by facilitating collaboration among satellites without the need to share the raw data. 
%Here, \cgls{fl} offers a solution for satellites to collaborate without the need to share the raw data.
Here, \cgls{fl} offers a solution by which satellites can collaboratively train an \cgls{ml} model without the need to share the raw data.
In canonical \cgls{fl} \cite{mcmahan2017communication}, which is a synchronous scheme, \cgls{ml} model parameters are communicated between the participating clients and \cgls{ps} in several iterations. 
However, when the clients are satellites, these iterations take longer time due to lack of consistent connection caused by non-visibility periods between the satellites and the \cgls{ps}, located in a \cgls{gs}, challenging the adoption of \cgls{fl} in satellite constellations \cite{matthiesen2022federated}. 
The visibility status of a satellite to a \cgls{gs} forms a pattern called visibility pattern.
\begin{figure}
    \centering
    \includegraphics[scale=0.45]{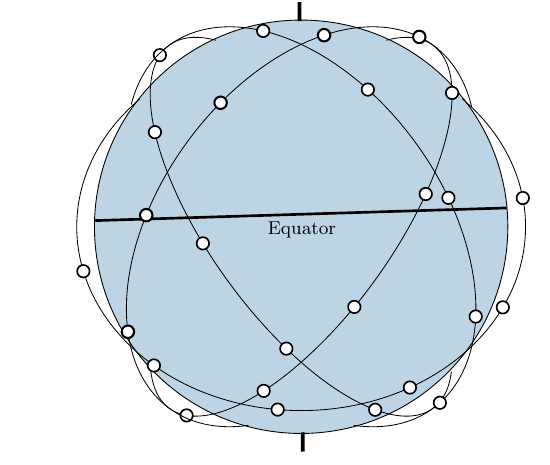}
    \caption{A satellite constellation with $P=5$ orbital planes and $K=40$ satellites. The number of equidistant satellites in all orbits is $K_p=8$, $\forall p \in \mathcal P = \{ 1, \dots, 5 \}$.}
    \label{fig: system_model}
\end{figure}
To deal with the non-visibility periods, an asynchronous \cgls{fl} algorithm is proposed in \cite{WCL_fedsat}, whose test accuracy is further improved in \cite{razmi2022scheduling}, exploiting the predictability of visibility patterns to schedule the transmissions between satellites and the \cgls{gs}.
Subsequently, to make also synchronous \cgls{fl} feasible for satellite constellations, an approach using intra-orbit \cglspl{isl} is presented in \cite{razmi2021board} to address the long delay caused by non-visibility periods.
This approach requires only one satellite from each orbit to be visible to the \cgls{gs} to access the necessary information from all other satellites in that orbit.
Nevertheless, in specific constellations, this approach may not effectively reduce the delay due to the prolonged non-visibility of all satellites in some orbits \cite{Razmi2023Journal}. Although this particular issue has not been studied in other subsequent works related to satellite \cgls{fl} such as \cite{elmahallawy2022fedhap, ostman2023decentralised, Chen2023,  so2022fedspace}, its effect can be mitigated by appropriate scheduling, addressed in the current study.
\begin{figure*}[!ht]
    \begin{minipage}[l]{1.0\columnwidth}
        \centering
        \includegraphics[scale=0.5]{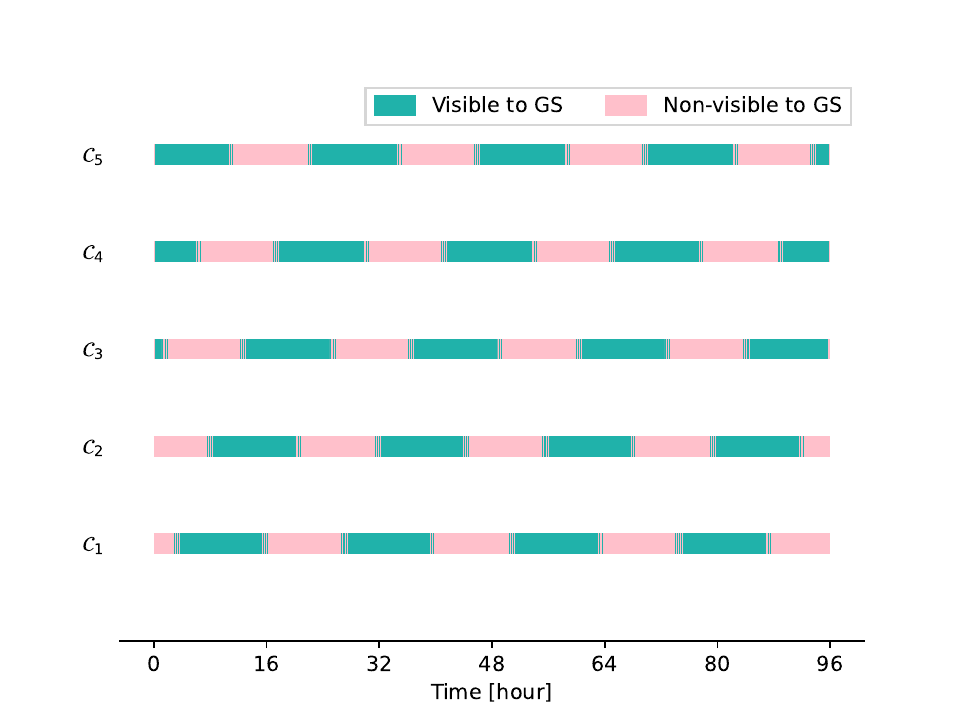}
        \subcaption{\footnotesize{GS located in Bremen, Germany}}\label{fig: Bremen}
    \end{minipage}
    \hfill{}
    \begin{minipage}[r]{1.0\columnwidth}
        \centering
        \includegraphics[scale=0.5]{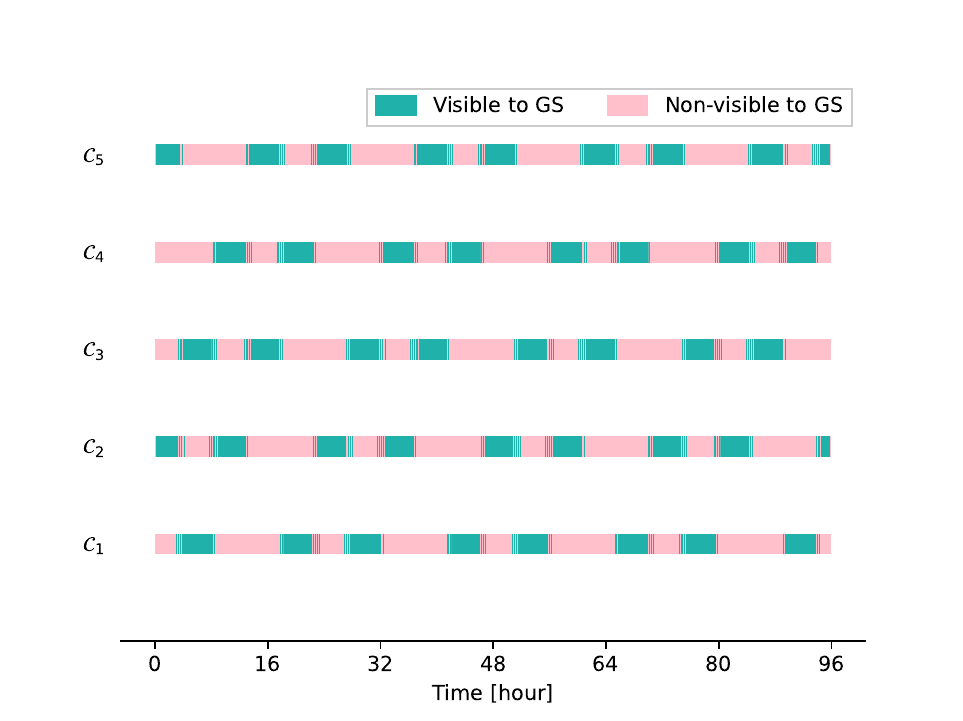}
        \subcaption{\footnotesize{GS located in S\~{a}o Paulo, Brazil}}\label{fig: Brazil}
    \end{minipage}
    \caption{Cumulative visibility pattern for a Walker Delta constellation with five orbits in an altitude of \num{2000} km, and inclination \ang{60}, each with eight equidistant satellites considering two \cgls{gs}s: a) located in Bremen, Germany, b) located in S\~{a}o Paulo, Brazil.}
    \label{fig: visibility pattern Bremen Brazil}
\end{figure*}

%Although this issue has not been studied in other subsequent works related to satellite \cgls{fl} such as \cite{elmahallawy2022fedhap, ostman2023decentralised, Chen2023,  so2022fedspace}, its effect can be mitigated by appropriate scheduling, addressed in the current study.
%In this paper, we extend our previous work \cite{razmi2021board, Razmi2023Journal} and propose a scheme that leverages both the predictability of the visibility pattern of each satellite and the cumulative visibility patterns of all satellites within an orbit. 
In this paper, as an extension to our previous works \cite{razmi2021board} and \cite{Razmi2023Journal}, we propose a scheduling scheme that leverages both the predictability of the visibility pattern of each satellite and the cumulative visibility pattern of all satellites within each orbit.
The satellites in each orbit are connected together in a ring formation through intra-orbit \cglspl{isl}. From the \cgls{gs} perspective, the connected satellites are considered as a unified entity, referred to as a cluster.
The proposed scheme comprises two schedulers: 1) \cgls{gu} scheduler and 2) \cgls{cu} scheduler. The \cgls{gu} scheduler, executed at the \cgls{gs}, determines the appropriate time instants for performing global iterations. On the other hand, \cgls{cu} scheduler is responsible to schedule \cgls{fl} procedures within each cluster based on the time instants received from the \cgls{gu} scheduler for global updates, allocating suitable learning duration to satellites within that cluster. The proposed schedulers lead to enhanced accuracy in a shorter duration.
%information

%%%%%%%%%%%%%%%%%%%
\section{System Model} \label{sec:System Model}

\subsection{Constellation Configuration}
We consider a satellite constellation as \cref{fig: system_model} with $P$ orbits and a \cgls{gs} located at a predetermined position on Earth. Each orbit $p \in \mathcal P = \{ 1, \dots, P \}$ is located at an altitude $h_p$ with inclination $a_p$ and contains $K_p$ equidistant satellites denoted by $\mathcal K_p = \{k_{p,1}, \dots, k_{p,K_p}\}$. The set of all satellites within the constellation is $\mathcal K = \bigcup_{p = 1}^P \mathcal K_p = \{ k_{1,1},\dots,k_{P,K_P} \}$ with the total number of $K = \sum_{p=1}^P K_p$ satellites.

\subsection{Cumulative Visibility Pattern}

%At any given time, satellites may be either visible or non-visible to the \cgls{gs}, depending on their respective locations. 
At any given time, a satellite in the orbit $p$ may be either visible or non-visible to the \cgls{gs}, depending on their respective locations.
The visibility periods are the intervals in which the connection between satellite and the \cgls{gs} is not obstructed by the Earth. In contrast, during non-visibility periods, Earth blocks the \cgls{los} link between them, resulting in intermittent connectivity. The visibility and non-visibility status of satellite with respect to time is referred to as visibility pattern of that satellite. Here, it is worth mentioning that due to the nature of satellite movements and Earth rotation, the visibility pattern of satellites in any constellation is known in advance.

Satellites in any orbit $p$ as are connected with intra-orbit \cglspl{isl} form a cluster, denoted by $\mathcal C_p$. Each satellite in $\mathcal C_p$ can exchange data, through other connected satellites, with the \cgls{gs} if there is at least one visible satellite to the \cgls{gs} in that cluster. 
In this regard, we introduce the concept of {\it{cumulative visibility pattern}} in which a cluster has a visible status if there is at least one visible satellite to the \cgls{gs} in that cluster.
%Therefore, it is reasonable to assume a satellite visible to the \cgls{gs} whenever there is at least one visible satellite in the cluster.

\cref{fig: visibility pattern Bremen Brazil} shows the cumulative visibility pattern with respect to wall-clock time for a constellation consisting of five orbits $P=5$, each orbit as a cluster containing eight satellites $K_p=8$, considering \cglspl{gs} located in two different positions: a) Bremen, Germany, b) S\~{a}o Paulo, Brazil. 
For the case \cgls{gs} located in Bremen, we observe that each cluster experiences a visibility period, followed by a prolonged non-visibility period. However, for the other case, with \cgls{gs} in S\~{a}o Paulo, the visibility and non-visibility periods are shorter but occur more frequently. 
These examples highlight the importance of cumulative visibility patterns in scheduling the procedures of \cgls{fl} for satellite constellations.

\subsection{Communication Model}

Each satellite can connect to the \cgls{gs} through \cgls{gsl} and can also communicate with the two nearest neighboring satellites in its orbit via intra-orbit \cglspl{isl}.
To facilitate these connections, each satellite is equipped with three antennas. The \cgls{gsl} antenna is directed toward the Earth’s center, while the intra-orbit \cgls{isl} antennas are positioned on both sides of the satellites, pointing toward the nearest neighboring satellite. 
Note that we only consider intra-orbit \cglspl{isl} since inter-orbit \cglspl{isl} require more complex control \cite{MatchingBeatriz19}.

Communication between two satellites $k$ and $i$ is only feasible when there is an unobstructed \cgls{los} link between them, meaning that their Euclidean distance, $d_{k, i}$, is less than the maximum slant range $d^S_{k, i} = \sqrt{h_{p(k)}^2+2r_Eh_{p(k)}} +\sqrt{{h_{p(i)}^2+2r_Eh_{p(i)}}}$, where $p(k)$ is the orbit index of $k$-th satellite, and $r_E$ is the Earth radius. Considering \cgls{gsl}, the connectivity between a satellite and the \cgls{gs} is feasible when $\frac{\pi}{2} - \angle (\vec r_{gs}, \vec r_k - \vec r_{gs}) \ge \alpha_e$, where $\vec r_{k}$ and $\vec r_{gs}$ denote the positions of satellite $k$ and the \cgls{gs} respectively, and $\alpha_e$ is the minimum elevation angle \cite{razmi2021board}. The maximum achievable data rate for \cgls{isl} transmission between satellites $k$ and $i$ is 
\begin{equation} \label{eq: rate}
    r_{k,i} = B \log_2 \left(1+\frac{P_t G_{k, i} G_{i,k} c^2}{ 16 \pi^2  d^2_{k,i} f_c^2 N_t}\right),
\end{equation}
where $P_t$ is the transmitted power, $G_{k,i}$ is the antenna gain of satellite $k$ toward satellite $i$, $c$ is the speed of light and $f_c$ is the carrier frequency. The total noise power is $N_t = k_B B T$ where $k_B = 1.380649 \times
10^{-23}\,\si{\joule/\kelvin}$ is the Boltzmann constant, $B$ and $T$ are the channel bandwidth and the receiver temperature respectively.

The time needed for satellite $k$ to deliver $S$ bits of data to satellite $i$ is obtained by summing up the transmission time and the propagation time as
\begin{equation} \label{eq: Time for transmitting}
    T_{k,i}^C= \frac{S}{r_{k,i}} + \frac{d_{k,i}}{\mathstrut c}.
\end{equation}
Similar to \cglspl{isl}, this model can also be applied to the \cglspl{gsl}. For simplicity, we consider the longest distance between each satellite and the \cgls{gs} within visibility period to derive the rate and propagation time.
 
\subsection{Learning Model}
Each satellite $k$ uses its local dataset $\mathcal D_k$ to train an \cgls{ml} model, consisting a set of parameters.
The overall objective is to learn the model parameters vector $\vec{w}$ that minimizes the global loss function $F(\vec{w})$ as 
\begin{equation} \label{eq: FL problem}
    F(\vec{w}) = \sum\nolimits_{k\in\mathcal K} \frac{D_k}{D} F_k(\vec{w})
\end{equation}
without sharing the local datasets with the \cgls{gs} or other satellites. The size of dataset $\mathcal D_k$ is $D_k$, and $D = \sum_{k\in\mathcal K} D_k$ is the total number of training samples. Moreover, the local loss function of satellite $k$ is 

\begin{equation} \label{eq: loss function of satellite k}
    F_k(\vec{w}) = \frac{1}{D_k} \sum\nolimits_{\vec{x}\in\mathcal D_k} f(\vec{x}, \vec{w}),
\end{equation}
where $f(\vec{x}, \vec{w})$ denotes the loss function for each sample $\vec{x}$ in the dataset. 
In overall $N$ global iterations, the \cgls{gs} cooperates with the satellites to minimize \cref{eq: FL problem} \cite{li2018federated}. In the $n$-th iteration, satellite $k$ first receives global model parameters $\vec{w}^{n-1}$ from the \cgls{gs}. Then, it performs $I$ local epochs of mini-batch \cgls{sgd} as \cref{alg:ssgd:ret} to minimize \cref{eq: loss function of satellite k}. 
 It is worth noting that satellites can solve $L^2$-regularization loss function $R_k(\vec{w}, \vec{w}^{n-1})=F_k(\vec{w})+ \frac{\lambda}{2} \norm{\vec{w} - \vec{w}^{n-1}}^2_2$, where $\lambda$ is the regularization parameter \cite{li2018federated}, instead of solving \cref{eq: loss function of satellite k}.
The local model parameters $\vec{w}_k^{n-1,I}$ are derived and sent back to the \cgls{gs} \cite{WCL_fedsat}. 
\begin{algorithm}[t!]
	\caption{Satellite Learning Procedure} \label{alg:Satellite SGD Procedure}
	\begin{algorithmic}[1]
		\Procedure{SatLearnProc}{$\vec{w}^{n-1}$}
			\State \textbf{initialize} $\vec{w}_k^{n-1,0} = \vec{w}^{n-1}, \quad i = 0$, \quad learning rate $\eta$ 
        \label{alg:ssgd:init}
			\For {$I$ epochs} \label{alg:ssgd:batchstart}
				\Comment $I$ epochs of mini-batch \cgls{sgd}
				\State $\tilde{\mathcal D}_k \gets $ Randomly shuffle $\mathcal D_k$
				\State $\mathscr B \gets $ Partition $\tilde{\mathcal D}_k$ into mini-batches of size $B$
				\For {each batch $\mathcal B\in\mathscr B$}
				\State $\vec{w}_k^{n-1,i} \gets \vec{w}_k^{n-1,i} - \frac{\eta}{|\mathcal B|} \nabla_{\vec{w}} \left(\sum_{\vec x\in\mathcal B} f(\vec x, \vec w) \right)$				
				\EndFor
                        \State $i \gets i + 1$
			\EndFor \label{alg:ssgd:batchend}
		      \label{alg:ssgd:compress}
			\State \Return $D_k \vec{w}^{n-1,I}_{k}$ \label{alg:ssgd:ret}
		\EndProcedure
	\end{algorithmic}
\end{algorithm}
Finally, the \cgls{gs} updates the global model parameters as
\begin{equation} \label{eq: global update}
    \vec{w}^{n} = \sum_{k=1}^{K} \frac{D_k}{D} \vec{w}_{k}^{n-1,I}
\end{equation}
and transmits the updated parameters, i.e., $\vec{w}^{n}$, again to the satellites for the next iteration.

\section{Scheduling scheme} \label{sec:Algorithm Description}

As mentioned in the previous sections, intra-orbit \cglspl{isl} allow satellites to exchange model parameters with the \cgls{gs} through other connected satellites, provided that at least one satellite from their cluster maintains a communication link to the \cgls{gs}. However, this is not always the case as depicted in \cref{fig: Bremen}. 
As we can see, the clusters experience long periods of non-visibility, during which no satellite from those clusters are visible to the \cgls{gs}.
This becomes problematic, especially when some clusters have completed their training procedures and transmitted their aggregated local model parameters to the \cgls{gs}, while others are non-visible.
In such scenarios, the \cgls{gs} has to wait a long period to receive the aggregated local model parameters from all clusters; therefore, performing a global update faces a prolonged delay.
To address this issue, one straightforward, but effective, way is to notify the clusters about the timing of the global updates, deduced from the predictability of the cumulative visibility pattern. This will empower the clusters to adjust their allocated time for training appropriately.

\begin{figure}
     \centering
     \includegraphics[scale=0.4]{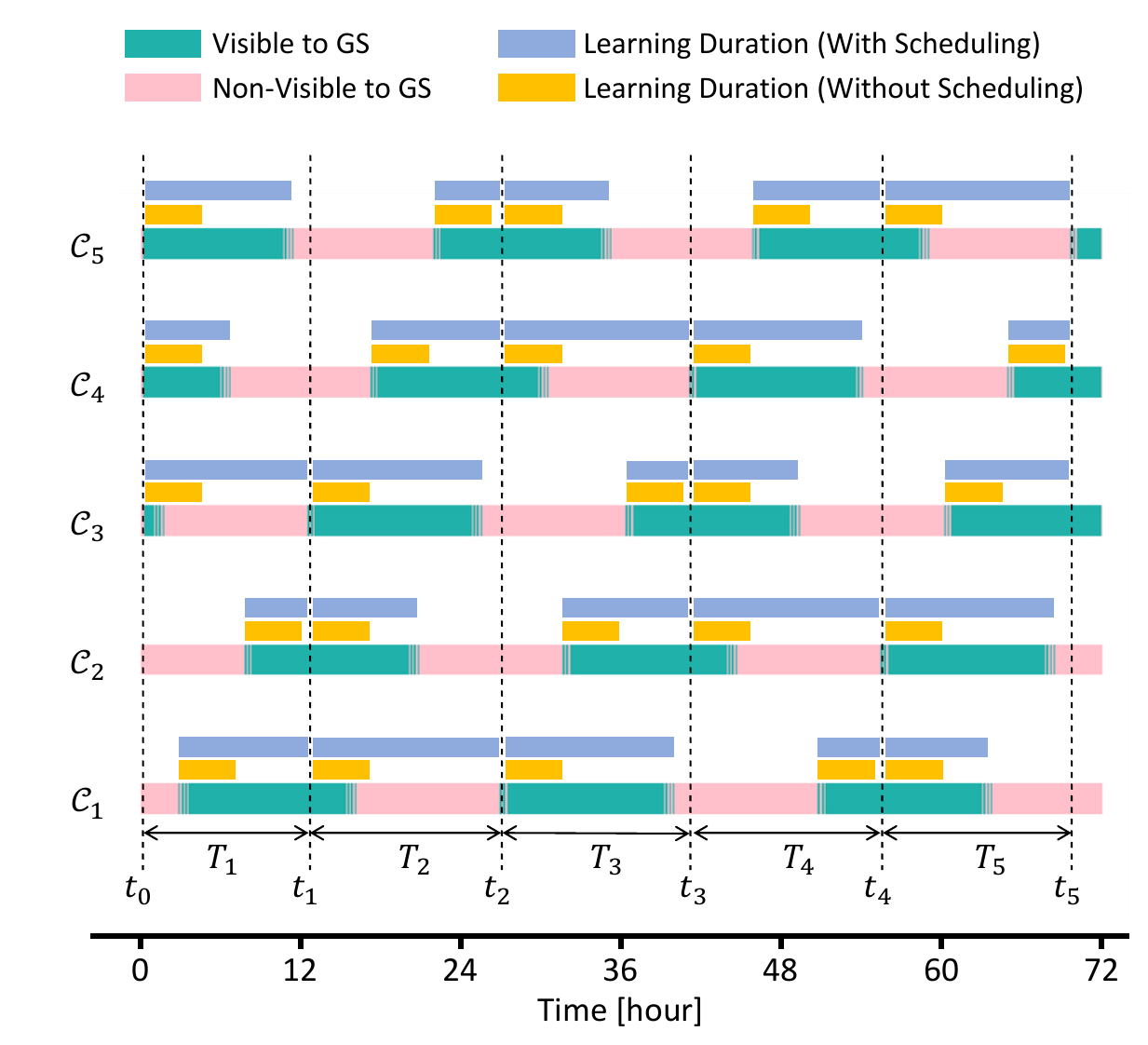}
     \caption{Comparison of the learning duration between the scheduled and unscheduled schemes in the Walker Delta constellation with five orbits, each containing eight satellites as considered in \cref{fig: Bremen}.} 
     \label{fig:Visiting_explain}
 \end{figure}

To this end, in this section, we introduce a scheduling scheme for satellite \cgls{fl}, comprising two schedulers: 1) a \cgls{gu} scheduler, and 2) a \cgls{cu} scheduler, as depicted in \cref{fig:scheduler}. 
The \cgls{gu} scheduler, executed at the \cgls{gs}, is responsible to determine the appropriate time instant for each global update in advance, denoted by $t_n$ for the $n$-th global update. Subsequently, the \cgls{gs} notifies the clusters about this determined instant. Having $t_n$, the \cgls{cu} scheduler, in each cluster $p$, derives the feasible duration for local training and accordingly adjusts the appropriate number of local epochs, $I_{n,p}$, that must be performed in the satellites in the $n$-th iteration. Let define {\it{time slot}} as the interval between the instants of two consecutive global updates, denoted by $T_n=t_n-t_{n-1}$ for the $n$-th global iteration. Consider time slots in the example presented in \cref{fig:Visiting_explain}. The figure shows that when the proposed scheduling scheme is applied, clusters, in each time slot, dynamically adjust their training duration, proportional to their visibility period. However, when this scheduling scheme is not applied, all clusters have to allocate a fixed duration for training in all time slots, leading to extended periods of inactivity in some time slots. 
In the following, we describe the \cgls{gu} and \cgls{cu} schedulers in detail.
\iffalse
\begin{figure}
    \centering
    \begin{tikzpicture}
         \draw[rounded corners, line width=1pt] (-1.5,0.5) rectangle (1.7,1.5);
         \draw (0.1,1) node { \begin{tabular}{c} Global Update (GU) \\  Scheduler \end{tabular}};

         \draw [rounded corners, line width=1pt] (3.9,0.5) rectangle (7.1,1.5);
         \draw (5.5,1) node {\begin{tabular}{c} Cluster Update (CU) \\ Scheduler \end{tabular}};

         \draw [arrows = {-Stealth[scale width=2]}, line width=1pt] (1.7,1) -- (3.9,1);
          \draw (2.7,1.3) node {($t_{n}, \vec{w}^{n-1}$)};

    \end{tikzpicture}
    %\caption{Global update (GU) scheduler and cluster update (CU) scheduler in the $n$-th iteration. The time instant for the $n$-th global iteration, $t_{n}$, and global model parameters, $\vec{w}^{n-1}$, are transmitted to each cluster.}
    \caption{The time instant for the $n$-th global iteration, $t_{n}$, is calculated by the \cgls{gu} scheduler and along with the global model parameters, $\vec{w}^{n-1}$, are transmitted to the \cgls{cu} schedulers.}
    \label{fig:scheduler}
\end{figure}
\fi

\begin{figure}
    \centering
    \includegraphics[scale=1]{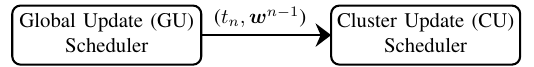}
    %\caption{Global update (GU) scheduler and cluster update (CU) scheduler in the $n$-th iteration. The time instant for the $n$-th global iteration, $t_{n}$, and global model parameters, $\vec{w}^{n-1}$, are transmitted to each cluster.}
    \caption{The time instant for the $n$-th global iteration, $t_{n}$, is calculated by the \cgls{gu} scheduler and along with the global model parameters, $\vec{w}^{n-1}$, are transmitted to the \cgls{cu} schedulers.}
    \label{fig:scheduler}
\end{figure}

\subsection{Global Update (GU) Scheduler}
The role of GU scheduler is to identify the earliest feasible time instant, $t_n$, at which $n$-th global update can be executed at the \cgls{gs}. This means all aggregated local model parameters for time slot $n$, $\vec{w}_{p}^{n} = \sum_{k=1}^{K_p} \frac{D_k}{D_p} \vec{w}_{k}^{n-1,I}$ where $D_p=\sum_{k=1}^{K_p} D_k$, from any cluster $p$ should be received at the \cgls{gs} by $t_n$. To this end, the \cgls{gu} scheduler uses the predictability of cumulative visibility pattern on one hand, and the required time to provide the aggregated local model parameters in the clusters on the other hand.
The \cgls{gu} scheduler, before the start of the $n$-th time slot, derives $t_n$ as
\begin{equation} \label{Eq: global update time instant}
    t_{n} = \max_{p}\enskip t_{n,p}^F,
\end{equation}
where $p\in \{1,\cdots,P \}$ and $t_{n,p}^F$ is the earliest feasible time instant by which the aggregated local model parameters of cluster $p$, $\vec{w}_{p}^{n}$, can be received at the \cgls{gs}. To formulate $t_{n,p}^F$, at first, the \cgls{gu} scheduler calculates a time instant called {\it{demand time}}, $t_{n,p}^{D}$, for time slot $n$ and cluster $p$.
It is the time instant by which the following procedures within the cluster $p$ are completed: 1) the cluster receives the global model parameters from the \cgls{gs} which demands, as \cref{eq: Time for transmitting}, a time interval of $T_{gs,p}^C$, 2) the global model parameters are distributed among satellites in the cluster through intra-orbit \cglspl{isl}, demanding time interval of ${T_{p}^{ISL}}$, 3) local training is performed in on-board of the satellites which demands a minimum learning duration of $T_{p}^L$, 4) the local model parameters are collected and aggregated within the cluster, again through intra-orbit \cglspl{isl}, demanding time interval of ${T_{p}^{ISL}}$. Hence, the demand time instant, $t_{n,p}^{D}$, is formulated as 

\begin{equation} \label{Eq: Demand time}
     t_{n,p}^{D} = t_{n,p,1}^R + T_{gs,p}^C + {T_{p}^{ISL}} + T_{p}^L + {T_{p}^{ISL}},
 \end{equation}
where $t_{n,p,1}^R$ is the first rise time, in the time slot $n$, in which the \cgls{gs} sends the global model parameters, i.e. $\vec{w}^{n-1}$, to the cluster $p$. The $m$-th rise time in the time slot $n$, $t_{n,p,m}^R$, is the time instant in which the cluster $p$ for the $m$-th time becomes visible to the \cgls{gs}.
Here, it is worth mentioning that a close observation of the cumulative visibility pattern occasionally shows frequent transitions between visibility and non-visibility states within a short period of time, resulting in having multiple rise and set times within each time slot.

After the above-mentioned procedures, the aggregated local model parameters, $\vec{w}_{p}^{n}$, are ready to be transmitted back to the \cgls{gs} if it is feasible, i.e., at least one satellite from the cluster becomes visible to the \cgls{gs}.
Therefore, at time instant $t_{n,p}^{D}$, if the cluster $p$ is in visible state, then it can immediately transmit $\vec{w}_{p}^{n}$ to the \cgls{gs}, and if it is in non-visible state, then it should wait until the next earliest visibility for transmitting those parameters. Therefore, by considering these two cases, $t_{n,p}^F$ is formulated as
\begin{equation} \label{eq: earliest visibility time}
    t_{n,p}^{F} = \begin{cases}
        t_{n,p}^{D}+T_{p,gs}^{C}, & \text{if $t_{n,p,m}^R \leq  t_{n,p}^{D} \leq t_{n,p,m}^S$}\\
        t_{n,p,m^{\ast}}^R+T_{p,gs}^C, & \text{otherwise}
    \end{cases}
\end{equation}
%Furthermore, a time interval of $T_{p,gs}^{C}$ is required for transmitting the aggregated parameters from the cluster to the \cgls{gs}.
where $t_{n,p,m}^S$ is the $m$-th set time, the time instant in which the $m$-th visibility period of the cluster $p$ ends, in the time slot $n$. In \cref{eq: earliest visibility time}, $T_{p,gs}^{C}$ is the time interval required for transmitting the aggregated parameters from the cluster to the \cgls{gs}. Furthermore, $t_{n,p,m^{\ast}}^R$ is the first subsequent rise time after $t_{n,p}^D$, meaning $m^{*}$ is determined by 
\begin{equation}
    \begin{aligned}
m^{*} = \operatorname*{argmin}_{m} \enskip t_{n,p,m}^R \quad
\textrm{s.t.} \quad  t_{n,p,m}^R \geq t_{n,p}^{D}.
\end{aligned} 
\end{equation}
By calculating $t_{n,p}^{F}$ for all clusters, $t_n$ can be derived using \cref{Eq: global update time instant}, which is transmitted along with the global model parameters, $\vec{w}^{n-1}$, to the clusters by the \cgls{gs} at the beginning of the $n$-th time slot.

\subsection{Cluster Update (CU) Scheduler} \label{sec: Cluster update scheduler}

The role of \cgls{cu} scheduler in each cluster is to determine the maximum possible number of local epochs that can be performed by the satellites of that cluster within each time slot $n$. 
To this end, at first rise time in the $n$-th time slot, i.e. $t_{n,p,1}^R$,
the \cgls{gs} transmits $\vec{w}^{n-1}$ and next global update time instant, i.e. $t_n$, to the one visible satellite of the cluster $p$, referred to as {\it{source}}.
Then, the source uses ${t_{n}}$ and the visibility pattern of satellites in that cluster to calculate the available time, $T_{n,p}^{A}$, which is the maximum time duration that can be allocated before $t_n$ for local training and communication procedures in the cluster. The time allocated for the communication procedures is for distributing the global model parameters and collecting the updated local model parameters among satellites through intra-orbit \cglspl{isl}. After performing these procedures, one satellite from the cluster, referred to as {\it{sink}}, transmits $\vec{w}_{p}^{n}$ to the \cgls{gs} when it is in the visible state. Here, two cases are possible: 1) at time $t_n$, the cluster is visible to the \cgls{gs}, and 2) the cluster is not visible to the \cgls{gs} at that time, $t_n$. For the first case, the cluster immediately transmits the aggregated model parameters, $\vec{w}_{p}^{n}$, to the \cgls{gs} at $t_n$. However, for the second case, the $\vec{w}_{p}^{n}$ should be transmitted to the \cgls{gs} at the latest set time before $t_n$. These two cases are formulated as
\begin{equation} \label{eq: available time}
    T_{n,p}^{A} = \begin{cases}
        t_{n} - t_{n,p}^{X}, & \text{if cluster $p$ is visible at $t_n$}\\
        t_{n,p,m^{+}}^S- t_{n,p}^{X}, & \text{otherwise}
    \end{cases}
\end{equation}
where $t_{n,p}^{X}=t_{n,p,1}^R + T_{gs,p}^C $, is the time instant in which the source receives the global model parameters from the \cgls{gs}. Moreover, $t_{n,p,m^{+}}^S$ is the latest subsequent set time before $t_{n}$, meaning $m^{+}$ is determined by
\begin{equation}
    \begin{aligned} 
m^{+} = \operatorname*{argmax}_{m} \enskip   t_{n,p,m}^{S} \quad \textrm{s.t.} \quad  t_{n,p,m}^S \leq t_{n}.
 \end{aligned}
\end{equation}
Using the calculated available time, $T_{n,p}^{A}$, by the source satellite, maximum possible number of local epochs that can be performed by the satellites of the cluster $p$ within time slot $n$ is derived as
\begin{equation}
    I_{n,p} = \left\lfloor \frac{T_{n,p}^A-2T^{ISL}_{p}-T_{p,gs}^C}{T_p^E} \right\rfloor,
\end{equation}
where $T_p^E$ is the required time for one local epoch which we, for simplification, assume it identical for all satellites in the cluster $p$; however, extending it to the heterogeneous case is straightforward if required. Moreover, $\lfloor \cdot \rfloor$ is the floor function. The upper-bound estimate of $T^{ISL}_{p}=\left\lceil{\frac{K_p}{2}}\right\rceil T^C_{k,j}$ is considered, which is the required time duration for distributing the global model parameters among the satellites or, conversely, collecting the local model parameters from them within the cluster, and $\lceil \cdot \rceil$ is the ceiling function. Moreover, $T_{p,gs}^C$ is the time duration required for communicating the aggregated local model parameters from the cluster to the \cgls{gs}. Then, the source determines the sink, a satellite in the cluster which is visible to the \cgls{gs} at time $t_{n,p}^{X}+T_{n,p}^A$. 

After determining $I_{n,p}$ and the sink by \cgls{cu} scheduler, we adopt the same scheme described in \cite{razmi2021board} for distributing and collecting parameters through intra-orbit \cglspl{isl} within each cluster.
It means the source transmits $I_{n,p}$ and the sink index along with $\vec{w}^{n-1}$ to its two nearest neighbor satellites in both directions in the orbit via intra-orbit \cglspl{isl}. Then, each receiving satellite forwards these pieces of information to their nearest neighbor satellite in the opposite direction of reception, continuing process until all satellites in the ring cluster have received the information \cite{razmi2021board}. Each satellite, after forwarding the received global parameters to its neighbor, initiates its learning as \cref{alg:Satellite SGD Procedure} by setting the number of local epochs to $I_{n,p}$.

Following the algorithm described in \cite{razmi2021board}, the two satellites in the cluster located farthest from the sink, after completing their learning phase, start to transmit the updated local model parameters to their nearest neighboring satellite in direction of the shortest path to the sink. 
These receiving satellites aggregate the received parameters with their own local updated parameters. Subsequently, they transmit the aggregated model parameters to the next nearest neighboring satellite in the opposite direction of reception. This aggregation and transmission sequence continues until the sink receives the aggregated local parameters from both directions. Finally, the sink aggregates the received parameters with its own parameters, derives $\vec{w}_{p}^{n}$ and transmits it to the \cgls{gs}. The \cgls{gs}, after receiving the aggregated parameters from all clusters, calculates the global model parameters as $\vec{w}^{n}=\sum_{p=1}^{P} \frac{D_p}{D} \vec{w}_{p}^{n}$ and continues with the next iteration.
 
%%%%%%%%%%%%%%%%%%%%
\section{Performance Evaluation}

To evaluate the performance of the proposed scheduling scheme for satellite \cgls{fl}, we perform experiments on a Walker Delta constellation with 40 satellites distributed across five orbits, located at the altitude of 2000~\si{km}. The inclination is set to \ang{60}. The \cgls{gs} is situated in Bremen, Germany, with a minimum elevation angle of \ang{10}. For the communication links, we set $f_c = \SI{20}{\giga\Hz}$, $B = \SI{500}{\mega\Hz}$, $P_t = \SI{40}{\dBm}$, $T = \SI{354}{\kelvin}$, and the antenna gains are set to \SI{32.13}{\dBi} \cite{leyva2022ngso}.

A deep CNN with \num{122570} parameters \cite{fraboni2023general} is trained on the CIFAR-10 dataset, which contains \num{60000} color images with the size of $32 \times 32$, categorized in ten classes.
The dataset distributed among satellites in a non-\cgls{iid} (non-i.i.d) manner using the Dirichlet distribution with a concentration parameter of 0.5 \cite{yurochkin2019bayesian, wang2020federated}. 
The batch size is 10, and the learning rate $\eta$ and regularization parameter $\lambda$ are set to \num{0.1} and \num{0} respectively. The time to complete a local epoch is set to $1$ hour. Moreover, the minimum demanded learning duration is set as the time needed to complete one local epoch by the satellites, i.e. $T_{p}^L = T_{p}^E$.
The simulation is conducted using the FedML library \cite{he2020fedml}.

\begin{figure}
     \centering
     \includegraphics[scale=1]{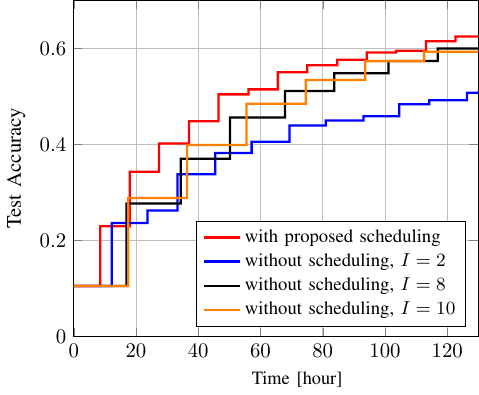}
     \caption{Test Accuracy of the proposed scheme with scheduling, with respect to wall-clock time, is compared with the scheme without scheduling for a Walker Delta constellation, comprising five orbits, each with eight satellites. In the scheme without scheduling, $I$ is the number of local epochs in each iteration and is fixed, while the scheme with scheduling dynamically adjusts this value based on the available time for each orbit in each time slot. }
     \label{fig:local epochs}
 \end{figure}

In \cref{fig:local epochs}, we compare the test accuracy between our proposed scheduling scheme and the scheme without scheduling in terms of wall-clock time. Unlike the proposed scheme which dynamically adjusts the number of local epochs, $I_{n,p}$, for each cluster $p$ and time slot $n$, the scheme without scheduling assigns a fixed number of local epochs, $I$, to satellites in all time slots. We evaluate three different cases for the scheme without scheduling: 1) two local epochs, i.e., $I=2$, 2) eight local epochs, i.e., $I=8$ and 3) ten local epochs, i.e., $I=10$.

The proposed scheme achieves higher accuracy compared to the scheme without scheduling for all the cases. 
Considering the case without scheduling with a lower $I$, e.g. $I=2$, 
the improvement achieved by the proposed scheme is attributed to the increased number of local epochs, $I_{n,p}$, in some clusters and time slots. The proposed scheme assigns $\{I_{1,1} = 5, I_{1,2} = 1, I_{1,3} =1, I_{1,4} =6, I_{1,5} =8\}$ to the clusters in the first time slot,
 and for the second time slot, it sets $\{I_{2,1} =7, I_{2,2} =9, I_{2,3} =5, I_{2,4} =1, I_{2,5} =2\}$.
This indicates the dynamic adjustment of $I_{n,p}$ for different $p$ and $n$ results in having higher number of local epochs than the fixed $I=2$ on average.

On the other hand, when considering the case without scheduling with a higher $I$, e.g. $I=10$, the improvement achieved by the proposed scheme is attributed to the increased number of global updates within given period of time. The proposed scheme adjusts $I_{n,p}$ to lower values for those clusters that become visible to the \cgls{gs} at a later time in each iteration. This prevents global updates from being delayed due to lack of local model parameters of those clusters, thus allowing for more frequent global updates.
For example, in the second time slot, as we mentioned above, $I_{2,4}=1$, allowing the forth cluster to transmit its local model parameters to the \cgls{gs} without long delay, despite becoming visible to the \cgls{gs} at a later time. However, in the scheme without scheduling, the absence of the dynamic adjustment results in the prolonged delays, as all clusters, even those experiencing delay, undergo the same high number of local epochs in all iterations.

%On the other hand, when considering the case without scheduling with a higher $I$, e.g. $I=10$, the improvement achieved by the proposed scheme is attributed to the increased number of global updates in given period of time. The proposed scheme adjusts $I_{n,p}$ to smaller values for those clusters that become visible to the \cgls{gs} at a later time in each iteration. This prevents any delays in global updates caused by a lack of local model parameters of those clusters, thereby enabling more frequent global updates.
%For example, as we mentioned above, in the second time slot, $I_{2,4}=1$, allows the forth cluster to transmits promptly its local model parameters to the \cgls{gs}, despite being the cluster that becomes visible to the \cgls{gs} late in this time slot. On the contrary, in the scheme without scheduling, the absence of the dynamic adjustment results in the prolonged delays, as all clusters undergo the same high number of local epochs in all iterations, even those delayed ones. 

%Another noteworthy observation from \cref{fig:local epochs} is that in the scheme without scheduling by comparing the cases with fixed low and high number of local epochs, i.e. $I=2$ and $I=10$, it is more advantageous to set $I$ to a fixed larger values.

%%%%%%%%%%%%%%%%%%%
\section{Conclusion}
We have presented a scheduling scheme to enable efficient federated learning in satellite constellations connected with intra-orbit inter-satellite links. 
Our approach leverages the predictability of visibility of each satellite as well as the cumulative visibility of all satellites in each orbit to effectively address intermittent connectivity challenges.
The proposed scheme incorporates two main schedulers: one for controlling the global update times and the other for managing learning procedures within each orbit. The scheduling scheme has enhanced the test accuracy by determining the appropriate time instants for global updates and dynamically adjusting the number of local epochs for each orbit and iteration, as confirmed by simulation results.

%%%%%%%%%%%%%%%%%%%
\bibliography{IEEEabrv,IEEEtrancfg,references.bib}

\balance
\end{document}